\documentclass[10pt,journal,compsoc]{IEEEtran}
\ifCLASSOPTIONcompsoc
  \usepackage[nocompress]{cite}
\else
  \usepackage{cite}
\fi

\ifCLASSINFOpdf
\else
\fi
\usepackage{float}
\usepackage{graphicx}
\usepackage{color}
\usepackage{psfrag}
\usepackage{subfigure}
\usepackage{algorithm}
\usepackage{algpseudocode}
\usepackage{graphics}
\usepackage{epsfig}
\usepackage{url}
\usepackage{amsthm}
\usepackage{amssymb, amsfonts, amsmath}
\usepackage{array}
\usepackage{changepage}

\usepackage{epstopdf}
\usepackage{graphicx}
\usepackage{epsfig}
\usepackage{enumerate}

\usepackage{amssymb}
\usepackage{amsmath}
\usepackage{graphicx}
\usepackage{epstopdf}
\usepackage{dcolumn}
\usepackage{bm}
\usepackage{color}
\usepackage{array}
\newcommand{\PreserveBackslash}[1]{\let\temp=\\#1\let\\=\temp}
\newcolumntype{C}[1]{>{\PreserveBackslash\centering}p{#1}}
\newcolumntype{R}[1]{>{\PreserveBackslash\raggedleft}p{#1}}
\newcolumntype{L}[1]{>{\PreserveBackslash\raggedright}p{#1}}

\begin{document}
%
\title{Identify Influential Spreaders in Asymmetrically Interacting Multiplex Networks}
%
%
%
%

\author{Qi Zeng,
        Ying Liu,
        Liming Pan,
        and Ming Tang
\IEEEcompsocitemizethanks{\IEEEcompsocthanksitem Q. Zeng and Y. Liu (corresponding author) are with the School of
Computer Science, Southwest Petroleum University, Chengdu 610500, China. Y. Liu is also with Big Data Research Center, University of Electronic Science and Technology of China, Chengdu 611731, China.\protect\\
E-mail: ying.liu@swpu.edu.cn
\IEEEcompsocthanksitem L. Pan is with the School of Computer Science and Technology, Nanjing Normal University, Nanjing 210023, China.\protect
\IEEEcompsocthanksitem M. Tang (corresponding author) is with School of Physics and Electronic Science, East China Normal University, Shanghai 200241, China ; Shanghai Key Laboratory of Multidimensional Information Processing, East China Normal University, Shanghai 200241, China.\protect\\
E-mail: tangminghan007@gmail.com
}
}

%
%

\markboth{Journal of \LaTeX\ Class Files,~Vol.~ , No.~ , }%
{Shell \MakeLowercase{\textit{et al.}}: Bare Demo of IEEEtran.cls for Computer Society Journals}
%


\IEEEtitleabstractindextext{%
\begin{abstract}
Identifying the most influential spreaders is important to understand and control the spreading process in a network. As many real-world complex systems can be modeled as multilayer networks, the question of identifying important nodes in multilayer network has attracted much attention. Existing studies focus on the multilayer network structure, while neglecting how the structural and dynamical coupling of multiple layers influence the dynamical importance of nodes in the network. Here we investigate on this question in an information-disease coupled spreading dynamics on multiplex networks. Firstly, we explicitly reveal that three interlayer coupling factors, which are the two-layer relative spreading speed, the interlayer coupling strength and the two-layer degree correlation, significantly impact the spreading influence of a node on the contact layer. The suppression effect from the information layer makes the structural centrality on the contact layer fail to predict the spreading influence of nodes in the multiplex network. Then by mapping the coevolving spreading dynamics into percolation process and using the message-passing approach, we propose a method to calculate the size of the disease outbreaks from a single seed node, which can be used to estimate the nodes' spreading influence in the coevolving dynamics. Our work provides insights on the importance of nodes in the multiplex network and gives a feasible framework to investigate influential spreaders in the asymmetrically coevolving dynamics.
\end{abstract}

}

\maketitle

\IEEEdisplaynontitleabstractindextext

%
\IEEEpeerreviewmaketitle

\IEEEraisesectionheading{\section{Introduction}\label{sec:introduction}}

%
%
%
%
\IEEEPARstart{M}{any} activities in society can be described as spreading processes on networks, such as the spreading of epidemic disease through contacts between human beings, information dissemination through email and mobile phone, and the diffusion of ideas among friends and community members. Identifying the most influential spreaders is an important step to control the spreading processes, e. g. to hinder epidemic outbreaks~\cite{freeman1978}, conduct successful advertisement for new products~\cite{christ2012} and protect key members in ecosystem~\cite{morone2018}. A commonly accepted way to rank and identify important nodes in a network is to use the centrality measures, such as degree, betweenness~\cite{freeman1977}, eigenvector centrality~\cite{bonacich2001}, PageRank~\cite{brin1998}, nonbacktracking centrality~\cite{martin2014}, and k-shell index~\cite{kitsak2010}. Based on the idea of centrality, there are a lot of achievements in the identification of important nodes in single layer networks~\cite{lv2016,Liu2017,makse2015}, which help us to better understand the network structure and function.

However in the real-world, from city infrastructure to human interaction patterns, many complex systems are interconnected and are better described by the multilayer network~\cite{mikko2014}. For example, the air transportation network can be described as a multilayer network where nodes represent airports and each commercial airline corresponds to a different layer~\cite{cardillo2013}. In social networks, individuals interact in different ways like being friends, colleagues, schoolmates, or interacting on different online social platforms~\cite{boccaletti2014}, where each type of connections is represented by a layer. The multiplex network is a particular type of multilayer network, in which a set of nodes represent the same individuals in all layers, and their edges in different layers represent various friendship patterns, such as the social networks.

It is thus natural to use the multilayer formalism to study the scenarios where different dynamical processes interplay, such as the cooperative contagion processes spreading in a host population~\cite{chen2017}, and the competitive epidemic spreading or opinion spreading on multilayer networks~\cite{sahneh2014,amato2017,wang2019}. A special case attracting much attention is the information-disease asymmetrically interacting processes~\cite{clara2013,clara2014}. When epidemic disease outbreaks in a district, the information on it is swiftly transmitted through the online social media, telephone, mass media, et al. The spreading of information suppresses the disease spreading because of the awareness people arise after receiving the information~\cite{funk2009}, and the spreading of disease promotes the diffusion of information, which are two asymmetrically interacting processes.

While many real-world complex systems can be modeled as multilayer networks, neglecting the multiple relationships between nodes or simply aggregating them into a single network alters the structural and dynamical properties of the system, leading to inaccurate identification of important nodes~\cite{battiston2014,manlio2016}. In recent years, there are some progress in identifying the critical nodes on multilayer network, which are in general extend the centrality measures from single network to multilayer network and the focus is on the multilayer structure alone. Examples are the Multiplex PageRank and eigenvector-based centrality in multiplex network. But how the interplay between the multiple processes on the multilayer network impact the functional importance of nodes is still unknown, and new ranking methods that considering the structural and dynamical interplay between layers are still lacking.

In modern times, the epidemic infectious disease spreads more easily due to the growing connectivity among metropolitan centers in the world urbanization progress~\cite{dirk2013}. Such as the ongoing coronavirus disease 2019 (COVID-19) has caused more than 20 millions confirmed cases and 700 thousands deaths in just a few months, impacting politically and economically on the daily lives of people. Timely and accurately identify the influential spreaders is thus of great importance to make optimal use of available resource to suppress epidemic disease~\cite{andrey2017}. In this manuscript, based on an asymmetrically interacting information-disease spreading model on multiplex network, we work on explicitly revealing how the structural and dynamical coupling of multiple layers influence the functional importance of nodes and then step further to propose a new method to rank and identify critical nodes in the asymmetrically interacting multiplex network. In our study, the multiplex network consists of two layers, where the spreading of information on one layer suppresses the spreading of disease on the other layer, while the spreading of disease promotes the information diffusion. As in real applications controlling the spreading of disease is usually the fundamental purpose, we concentrate on the influence of nodes in disease spreading.

Firstly, by taking the degree and eigenvector centrality of nodes on the physical contact layer as the benchmark measure, we study on how the two-layer coupling factors impact their accuracy to predict the spreading influence of nodes. As the spreading influence of nodes in the contact layer is suppressed by information spreading, the accuracy of centralities is impacted. The application of discovering the performance of centralities under different structural and dynamical parameters is two-fold. One the one hand, due to the difficulty in collecting network data, when we only have the contact layer data and obtain the centrality of nodes from the contact layer alone, we can evaluate how accurate the obtained centrality can predict the influential spreaders. On the other hand and more importantly, based on the network data of both layers, we should define new measure to accurately identify the most influential spreaders in the asymmetrically interacting dynamics.

Then we propose an effective framework to rank the node influence in the asymmetrically interacting multiplex networks. Specifically speaking, by mapping the coevolving spreading dynamics into bond percolation and using the message-passing approach, we calculate the spreading outbreak size for each node as seed, which can be used to rank the influence of disease spreaders in the multiplex network. The accurate identification of disease spreaders is very applicable in real-world epidemic control.

To the best of our knowledge, this work is the first step in explicitly studying how the two-layer coupling factors impact the spreading influence and the centrality of nodes in identifying influential spreaders, and ranking the node influence by considering the coupling factors. Our main contributions in this paper are as follows:
\begin{itemize}
\item{We discover how three coupling factors, which are the relative spreading speed of the two layers, the coupling strength and the inter-layer degree correlation, impact the accuracy of centrality measures in predicting spreading influence in multiplex network.}

\item{By mapping the coevolving spreading dynamics into percolation, we propose a method to accurately rank the spreading influence of nodes in the multiplex network.}

\item{We numerically evaluate our proposed method in multiplex networks and show its superiority over the benchmark centralities in identifying influential spreaders.}
\end{itemize}

The rest of the paper is organized as follows. In section 2, we briefly introduce related works. In section 3, the information-disease spreading model on multiplex network is described. In section 4, we demonstrate the impact of interlayer coupling factors on the accuracy of centralities. In section 5, we map the coevolving spreading dynamics into percolation and propose a new ranking method. Finally in section 7, we give the conclusions.

\section{Related works}
To identify important nodes in the multilayer network, a lot of centrality measures have been proposed, which are in general extended from centrality in single networks. For example, the Multiplex PageRank is a natural extension of PageRank in multiplex network, which considers the centrality of a node in one network is affected by the centrality of the node in another network~\cite{halu2013}. Another Functional Multiple PageRank is defined on the weight of multilinks (connections in different layers) where the link overlap between layers is considered~\cite{iaco2016}. The eigenvector centrality in multiplex networks takes into account the mutual influence of layers~\cite{sola2016}, and a supracentrality matrix which couples the centrality matrices of the individual layers is used ~\cite{taylor2019}. The tensor framework mathematically describes the intralayer and interlayer relationships~\cite{manlio2013}, and tensor decomposition is then used to identify critical nodes~\cite{wang2017}. In the tensorial formalism, Domenico et al. generalizes the eigenvector centrality, PageRank and betweenness to multilayer network and define the versatility to identify the most important nodes~\cite{domenico2015}. A family of multilayer PCI (Power Community Index) measures generalized from h-index considers the density of a node's intra and inter connections and can be computed from the local structure of the network~\cite{pavlos2019}.

Although there are such progresses on identifying critical nodes on multilayer network, the focus is on the structure of the network, while neglecting how the structural and dynamical coupling between multiple layers influence the dynamical importance of nodes in the network. In multilayer network, the interplay between the network structure and spreading dynamics on top of it will largely influence the role of nodes in the network~\cite{manlio2013centrality}. Identifying the critical nodes must take both the structural and dynamical characteristics of the network into consideration. A few works summed up the centrality of nodes in both layers and synthesized the dynamical parameters to form a measure to quantify the node influence in symmetrically interacting interconnected networks, and the results showed that taking both layers' structure and dynamical parameters into consideration leaded to a better ranking result of the critical nodes ~\cite{zhao2014}. How the coevolving dynamics impacts on the functional importance of nodes and accurate identification of important nodes remains an open question.

\section{The information-disease spreading model on multiplex network}
\begin{figure*}[htbp]
\begin{center}
\includegraphics[width=15 cm]{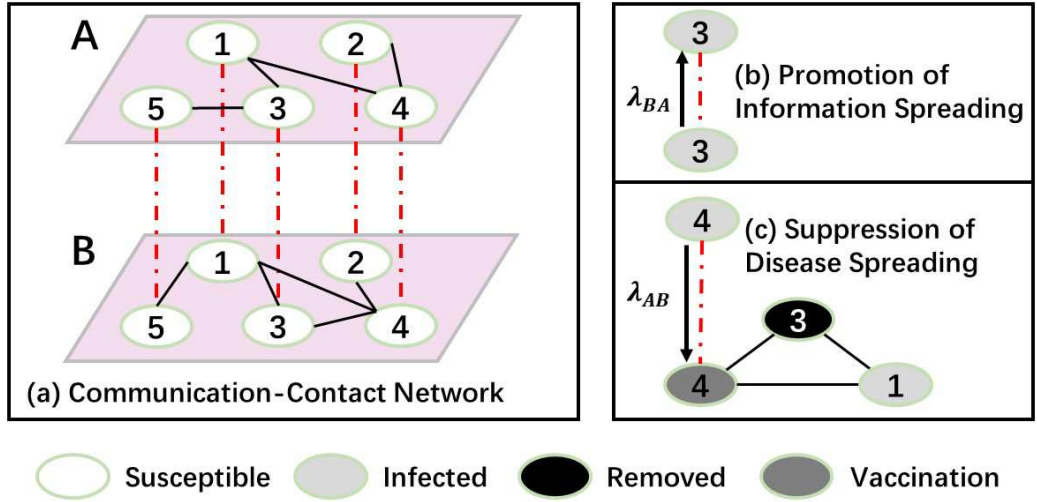}
\caption{The information-disease spreading processes on a communication-contact multiplex network. (a) The multiplex network consists of a communication layer A with SIR dynamics and a physical contact layer B with SIRV dynamics. (b) An infected node in layer B promotes the information spreading by changing its counterpart into informed state with rate $\lambda_{BA}$. (c) An informed node in layer A suppresses the disease spreading by changing its counterpart into vaccinated state with rate $\lambda_{AB}$.}
\label{figure1}
\end{center}
\end{figure*}

We use an asymmetrically interacting model to describe the coevolving dynamics of disease and information spreading~\cite{wang2014, wang2016} on multiplex network as shown in Fig.~\ref{figure1}. Consider a multiplex network consisting of two layers. The upper layer represents the information communication network labeled as layer A and the bottom layer represents the physical contact network labeled as layer B. In the communication layer (layer A), the classical susceptible-infected-recovered (SIR) model is used to describe the spreading of information on disease. In the SIR model, node can be in one of the three states: (1) susceptible (S), in which the individual has not received any information about the disease, (2) infected (or informed for information transmission I), in which the individual is aware of the disease and is able to transmit the information, or (3) recovered (R) in which the individual has received the information but is not willing to transmit the information to others. The informed node tries to transmit information to its neighbors at rate $\beta_A$. The informed node recovers with rate $\mu_A$ in the next time step. Once a node becomes recovered, it will remain in the state in the subsequent time steps.

The spreading of disease on the physical contact layer B is described by the SIRV model~\cite{ruan2012}, where a vaccinated(V) state is introduced. A vaccinated node will not be infected by any node. The SIR dynamics is the same as layer A, with the infection rate and recovery rate denoted as $\beta_B$ and $\mu_B$ respectively.
The dynamical coupling of the two layers is as follows. For an informed individual in layer A, if its counterpart in layer B is susceptible, then the counterpart node changes to vaccination state with rate $\lambda_{AB}$. The parameter $\lambda_{AB}$ ranges in $[0,1]$ to represent the extent to which people take care of the information and are willing to take vaccination. For a susceptible node in layer B, if its counterpart in layer A is in susceptible state, then in the next time step it may become either vaccinated with rate $\lambda_{AB}$ if its counterpart in layer A is getting informed or infected by its infected neighbors in layer B. In this case, vaccination and infection will compete for the chance to affect such susceptible node in layer B. Take $p_A$ and $p_B$ as the probability that the vaccination or infection wins respectively, then
\begin{equation}\label{pa}
p_A=\frac{1-(1-\beta_A)^{n_i^A}}{[1-(1-\beta_A)^{n_i^A}]+[1-(1-\beta_B)^{n_i^B}]}
\end{equation}
and
\begin{equation}\label{pb}
p_B=\frac{1-(1-\beta_B)^{n_i^B}}{[1-(1-\beta_A)^{n_i^A}]+[1-(1-\beta_B)^{n_i^B}]},
\end{equation}
 where $n_i^A$ is the number of informed neighbors of the counterpart node in layer A, and $n_i^B$ is the number of infected neighbors of the considered node in layer B. If vaccination wins out, the node in layer A is informed with probability $1-(1-\beta_A)^{n_i^A}$ and then the couterpart node in layer B is vaccinated with rate $\lambda_{AB}$. Else, if the infection wins out, the node is infected in layer B with probability $1-(1-\beta_B)^{n_i^B}$ and its counterpart node changes to the informed state with rate $\lambda_{BA}$, which ranges in $[0,1]$. This represents the extent to which an infected individual is aware of and willing to transmit the information of the epidemic disease.
\section{Revealing the impact of interlayer coupling factors}
It is pointed out that the asymmetrically interacting dynamics will alter the activities of nodes on the multiplex networks~\cite{clara2013}. Now we study on how three dynamical and structural two-layer coupling factors, which are the the relative spreading speed of the two layers, the dynamical coupling strength and the interlayer degree correlation, affect the spreading influence of nodes and thus change the accuracy of centrality in predicting their influence in the physical contact layer, which is our focus.
The relative spreading speed of the two layers is defined as
\begin{equation}\label{r}
\gamma^{\lambda}_{AB}=\lambda_A/\lambda_B,
\end{equation}
where $\lambda_A=\beta_A/\mu_A$ and $\lambda_B=\beta_B/\mu_B$ are the effective transmission rate in layer A and B respectively. For simplicity, we set $\mu_A=\mu_B=1$. The dynamical coupling strength of the two layers is represented by the two parameters $\lambda_{AB}$ and $\lambda_{BA}$, which are the vaccination rate and informed rate respectively. The interlayer degree correlation $m_s$ is quantified by the Spearman rank correlation coefficient, which is defined as
\begin{equation}\label{ms}
m_s=1-6\frac{\sum_{i=1}^N\Delta_i^2}{N(N^2-1)},
\end{equation}
where $N$ is the network size and $\Delta_i$ is the rank difference of node $i$ in the degree sorting list in each layer.
\subsection{Construct the multiplex network}
In simulations, we use the uncorrelated configuration model (UCM) to generate each layer of the multiplex network, which follows a power law degree distribution $p(k)\sim k^{-\gamma}$. We first construct layer A with $N=10000$, the power exponent $\gamma=2.6$, and average degree $<k>=6$. The minimal degree $k_{min}=3$, and the maximal degree $k_{max}=\sqrt{N}$. Then we copy the nodes of layer A, randomly exchange their degree sequence and generate the edges to form layer B. Each node in layer A has a counterpart in layer B. At the beginning of the interacting dynamical processes, all nodes are set to be susceptible in both layers except the seed node. The seed node is infected in layer B and its counterpart in layer A is informed, which will initiate the disease spreading on layer B and information spreading on layer A respectively. The spreading processes stop until on both layers there is no infected (informed) nodes. We record the final proportion of recovered nodes in layer B as the spreading influence of the seed node, which is averaged over 100 independent realizations.
\subsection{The benchmark degree and eigenvector centrality}
Suppose an undirected network is represented as G(V, E), where $V=\{v_1, v_2,..., v_n\}$ is the set of nodes and $E=\{e_1, e_2,..., e_m\}$ is the set of edges. The adjacency matrix of the graph G is $A_{n*n}=a_{ij}$, where $a_{ij}=1$ means there is an edge between node $i$ and $j$, otherwise $a_{ij}=0$. The degree $k_i=\sum_{1}^na_{ij}$ of node $i$ is defined as the number of its direct neighbors, which is a simple but effective way to quantify the potential influence of nodes in the network. The larger the degree, the more neighbors the node is able to influence directly. The time complexity of calculating degree is $O(N)$, where $N$ is the size of the network.

The idea of eigenvector centrality is that not all neighbors are equivalent. The node is important if it connects to many neighbors which are themselves important. The eigenvector centrality of node $i$ is defined as
\begin{equation}\label{eigenvector}
e_i=\lambda^{-1}\sum_{n=1}^Na_{ij}e_j,
\end{equation}
where $\lambda$ is the largest eigenvalue of the adjacency matrix A, $e=\{e_1, e_2, ..., e_n\}^T$ is the eigenvector of matrix A corresponding to the largest eigenvalue $\lambda$. If writing in the form of matrix, then it is $\lambda e=Ae$.
In our work, the degree and eigenvector centrality of nodes in layer B are used as benchmark methods to identify the most influential nodes in disease spreading.
\subsection{Evaluation methods}
We use the Kendall's tau correlation coefficient~\cite{kendall1938} and the imprecision function~\cite{kitsak2010} to quantify how accurate the considered measures can predict the disease-spreading influence of nodes in the asymmetrically interacting processes. The Kendall's tau correlation coefficient quantifies the consistency of two ranking lists for a set of objects. It is defined as
\begin{equation}
\tau=\frac{\sum_{i<j}sgn[(x_{i}-x_{j})(y_{i}-y_{j})]}{\frac{1}{2}n(n-1)},
\end{equation}
where sgn(x) is a sign function, which returns 1 if $x>0$, -1 if $x<0$, and 0 if $x=0$. $n$ is the number of nodes in the lists. $x_{i}$ and $x_{j}$ are the rank of nodes $i$ and $j$ in ranking list 1, while $y_{i}$ and $y_{j}$ are the rank of nodes $i$ and $j$ in ranking list 2. If the node pair $i$ and $j$ has a concordant order in ranking list 1 and 2, $(x_{i}-x_{j})(y_{i}-y_{j})>0$. If the node pair $i$ and $j$ has a discordant order in ranking list 1 and 2, $(x_{i}-x_{j})(y_{i}-y_{j})<0$. If the nodes $i$ and $j$ has an identical rank in either list 1 or list 2, $(x_{i}-x_{j})(y_{i}-y_{j})=0$. In our applications, nodes are ranked by centrality measure in ranking list 1 and are ranked by their real spreading influence in ranking list 2. A large correlation coefficient $\tau$ implies that the centrality can better predict the spreading influence of nodes.

The imprecision function is defined as
\begin{equation}\label{imprecision}
\varepsilon(p)=1-\frac{M(p)}{M_{eff}(p)},
\end{equation}
where $p$ is the fraction of the network size $N$ ($p\in[0,1]$). $M(p)$ is the average spreading influence of $pN$ nodes with the highest centrality, and $M_{eff}(p)$ is the average spreading influence of $pN$ nodes with the highest spreading influence. This function quantifies how close to the optimal spreading is the average spreading of the $pN$ nodes with the highest centrality. The smaller the $\varepsilon$ value, the more accurate the centrality is a measure to identify the most influential spreaders.
\subsection{Results}
\begin{figure*}[htbp]
\begin{center}
\includegraphics[width=15 cm]{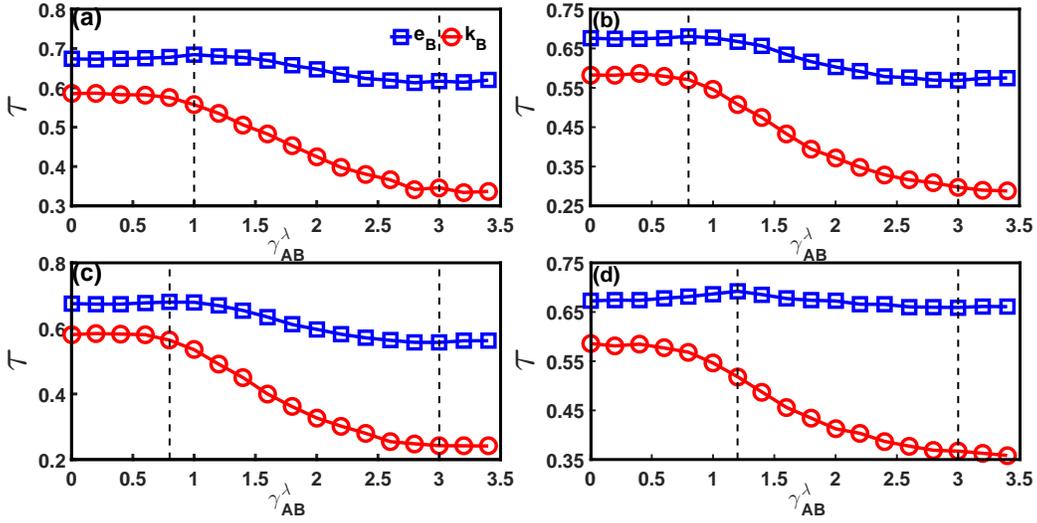}
\caption{The $\tau_{e_B}$ and $\tau_{k_B}$ as a function of $\gamma^{\lambda}_{AB}$. The other parameters are set as (a) $\lambda_{AB}=0.7$, $\lambda_{BA}=0.1$, $m_s=0.5$, (b) $\lambda_{AB}=1.0$, $\lambda_{BA}=0.1$, $m_s=0.5$, (c) $\lambda_{AB}=1.0$, $\lambda_{BA}=0.1$, $m_s=0.7$ and (d) $\lambda_{AB}=1.0$, $\lambda_{BA}=0.3$, $m_s=0.5$.}
\label{figure2}
\end{center}
\end{figure*}

Firstly, we study on the impact of relative spreading speed of two layers on the accuracy of centrality in layer B. The accuracy of centrality in predicting node influence is quantified by the Kendall's $\tau$ correlation of centrality and spreading influence as shown in Fig.~\ref{figure2}. To concentrate on the relative spreading speed $\gamma_{AB}^{\lambda}$, we fix other coupling parameters and demonstrate the results. The impact of these parameters when they vary will be discussed in later part. At $\gamma_{AB}^{\lambda}=0$, it corresponds to the case when there is only disease spreading on layer B. The change of correlation $\tau$ is due to the change of spreading influence of nodes, which is suppressed by information spreading in layer A.
It can be seen from Fig.~\ref{figure2} that $\tau$ decreases with $\gamma^{\lambda}_{AB}$. In Fig.~\ref{figure2} (a) when $\gamma^{\lambda}_{AB}<1.0$, the $\tau_{e_B}$ and $\tau_{k_B}$ are relatively stable. This means when the information spreads slowly, or even slower than the disease, it has little impact on the disease spreading. When $\gamma^{\lambda}_{AB}$ increases to the range [1.0, 3.0], $\tau_{e_B}$ and $\tau_{k_B}$ largely decrease. This is because when the information spreads faster than the disease, more nodes in layer B will be vaccinated. The spreading influence of nodes in layer B is suppressed by the information spreading. The faster the information spreads, the more nodes get vaccinated, making the $e_B$ and $k_B$ less accurate. Consider the real-world scenario that the epidemics control agency wants to identify the most influential disease spreaders and quarantine them. When the information spreads slowly, it is workable to identify the influential spreaders from the structure of contact network. But when information spreads fast, using only the contact data is not adequate any more. As for $\gamma^{\lambda}_{AB}>3.0$, where the information spreads even faster, the value of $\tau$ becomes stable. This is because when the information spreads very fast, the number of vaccinated nodes achieves its upper limit, and the spreading influence of nodes impacted by the vaccination will not change. The change of other three parameters $\lambda_{AB}$, $\lambda_{BA}$ and $m_s$, do not influence the decreasing trend of $\tau$, as demonstrated in Fig.~\ref{figure2} (b)-(d).

\begin{figure*}[htbp]
\begin{center}
\includegraphics[width=15 cm]{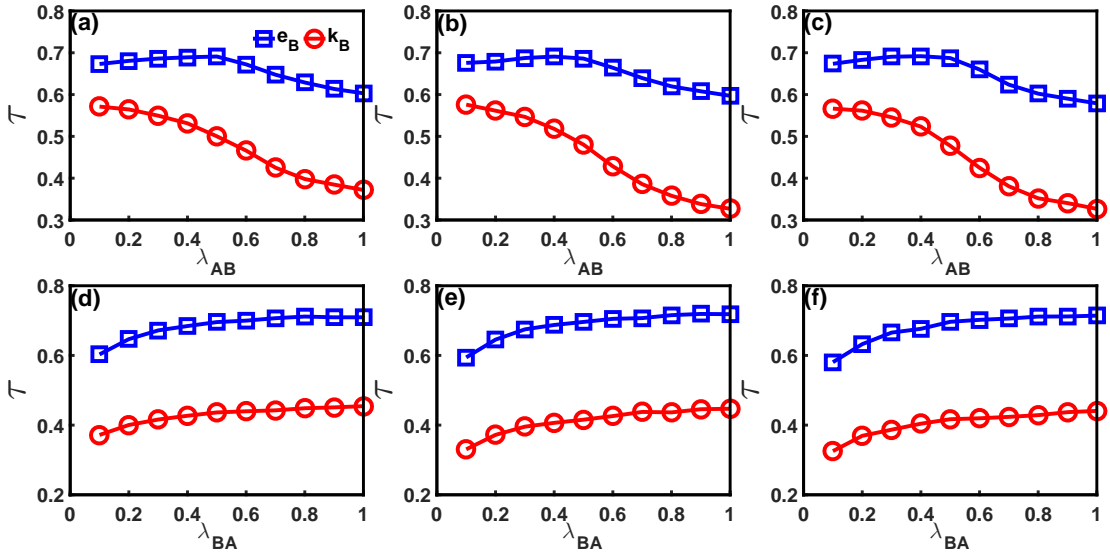}
\caption{The $\tau_{e_B}$ and $\tau_{k_B}$ as a function of $\lambda_{AB}$ and $\lambda_{BA}$ respectively. The other parameters are set as (a) $\gamma^{\lambda}_{AB}=2.0$, $m_s=0.5$, $\lambda_{BA}=0.1$, (b) $\gamma^{\lambda}_{AB}=2.0$, $m_s=0.7$, $\lambda_{BA}=0.1$, (c) $\gamma^{\lambda}_{AB}=2.4$, $m_s=0.5$, $\lambda_{BA}=0.1$,(d) $\gamma^{\lambda}_{AB}=2.0$, $m_s=0.5$, $\lambda_{AB}=1.0$, (e) $\gamma^{\lambda}_{AB}=2.0$, $m_s=0.7$, $\lambda_{AB}=1.0$ and (f) $\gamma^{\lambda}_{AB}=2.4$, $m_s=0.5$, $\lambda_{AB}=1.0$.}
\label{figure3}
\end{center}
\end{figure*}

Next, we work on the interlayer coupling strength. There are two parameters $\lambda_{AB}$ and $\lambda_{BA}$ reflecting the coupling strength between two layers. From Fig.~\ref{figure3}(a), we can see that with the increase of $\lambda_{AB}$, $\tau_{k_B}$ decreases significantly, which implies that the simple degree centrality in layer B becomes worse to rank the node influence. As for $\tau_{e_B}$, it first increases a little and then decreases. In general, $\tau$ decreases with the increase of $\lambda_{AB}$. This is because when $\lambda_{AB}$ increases, the effect of layer A on B are getting more strong, thus the centrality on layer B are becoming less accurate under the asymmetrically interacting processes. The Fig.~\ref{figure3} (b) and (c) displays similar trends as (a). In Fig.~\ref{figure3}(b), it can be seen that with the increase of $\lambda_{BA}$, $\tau$ increases. When $\lambda_{BA}$ is small, the amount of informed nodes in layer A caused by the notification from layer B is small. So the spreading of information relies more on the structure and centrality of nodes in layer A. In this case, the suppression of disease is more dependent on the structure of layer A, leading to the relatively low accuracy of centrality in layer B to predict the spreading influence of nodes. When $\lambda_{BA}$ becomes large, more informed nodes in layer A are caused by the notification from their counterparts in layer B, so the the distribution of informed nodes is more random and the suppression of disease then depends less on the centrality of nodes in layer A. When the number of informed nodes is large enough, the distribution of them can be considered as uniform in layer A. In this case, the spreading influence of nodes in layer B are reduced proportionally to their degree. The larger $\lambda_{BA}$ is, the stronger such effect is. Thus the $\tau$ increases as $\lambda_{BA}$ increases. The Fig.~\ref{figure3} (e) and (f) displays similar trends as (d).In real-world scenarios, if the infected individuals can timely report their heathy status, corresponding to large $\lambda_{BA}$, then it is easier to identify correctly the influential disease spreaders. Otherwise, as the infected ones are hidden, it becomes more difficult to identify the influential spreaders under the interplay between information spreading and disease spreading. This will prevent effective epidemic control from the healthy agencies.

\begin{figure*}[htbp]
\begin{center}
\includegraphics[width=15 cm]{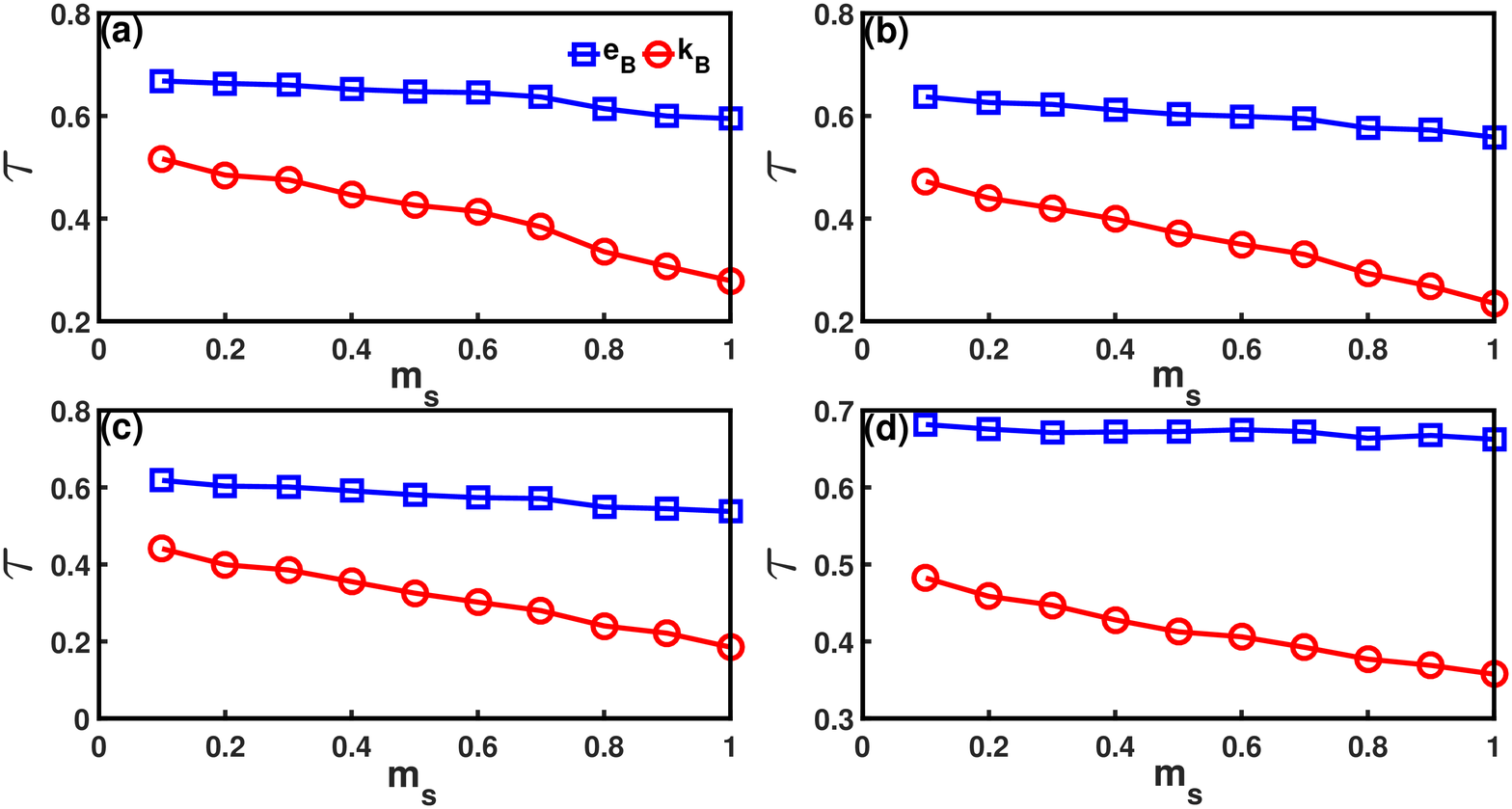}
\caption{$\tau_{e_B}$ and $\tau_{k_B}$ as a function of $m_s$. The other parameters are set as (a) $\gamma^{\lambda}_{AB}=2.0$, $\lambda_{AB}=0.7$, $\lambda_{BA}=0.1$, (b) $\gamma^{\lambda}_{AB}=2.0$, $\lambda_{AB}=1.0$, $\lambda_{BA}=0.1$, (c) $\gamma^{\lambda}_{AB}=2.4$, $\lambda_{AB}=1.0$, $\lambda_{BA}=0.1$ and (d) $\gamma^{\lambda}_{AB}=2.0$, $\lambda_{AB}=1.0$, $\lambda_{BA}=0.3$.}
\label{figure4}
\end{center}
\end{figure*}

Finally, we discuss the effect of degree correlation $m_s$ between layers. The spearman rank correlation coefficient is used to quantify the degree correlation of nodes in two layers. As shown in Fig.~\ref{figure4}, the $\tau$ decreases with the increase of degree correlation $m_s$. Remember that at the initial step of interacting spreading, a seed node in layer B is infected  and its counterpart in layer A is informed. The affected range of disease spreading from the seed node is determined by the centrality of seed in both layer A and B. From the respect of degree correlation, all seeds can be divided into four cases: (1) nodes with large $k_A$ and small $k_B$. In this case, because of their small degree in layer B, the spreading influence of such nodes is relatively small. Although the information spreading is large due to their large $k_A$, the suppression effect is less obvious. Thus the performance of centrality is not largely affected. (2) nodes with small $k_A$ and large $k_B$. In this case, due to the centrality of seed, the disease spreading is relatively large and the information spreading is small, thus there will be a small number of vaccinated nodes and the suppression effect is as well small. (3) nodes with large $k_A$ and large $k_B$. For these nodes as initial spreaders in both layers, the suppression effect for disease spreading is obvious, making the centrality in predicting the disease spreading less accurate. The larger degree correlation $m_s$ is, the more such kind of nodes are in the network, corresponding to the largely reduced $\tau$ in Fig.~\ref{figure4}. (4) nodes with small $k_A$ and small $k_B$. In this case, neither the disease nor information will break out and these nodes are ranked low in the list. In all, when the two-layer degree correlation is large, the suppression effect on the disease spreading is the largest, and the $\tau$ of centrality and spreading influence is impacted the most.
The above results implies that in real-world epidemic control where the information spreading and disease spreading interplays, using the contact network data is not adequate to accurately identify the critical spreaders, especially when the information spreads fast, the vaccination willingness of people is strong and the degree correlation is large. So to identify the influential spreaders more accurately in multiplex network, we need new framework and method which is our work in the next part.
\section{Mapping to bond percolation}
In this part, we map the coevolving spreading dynamics into bond percolation and calculate analytically the prevalence when the epidemic is originated from a single seed $i$ on the multiplex network by using the message passage method. The message passing approach is an inference method that can provide exact predictions or good approximations analytically in many problems in network sciences, such as predicting the size of the giant component in percolation~\cite{allard2019}. In the SIR dynamics, the ultimate outbreak size corresponds to the size of the giant component in percolation, which is our interests.

We first introduce the percolation on a single network. In the percolation process, edges are occupied with probability $T_p$ called transmissibility, and a giant component appears if $T_p$ is sufficient high. The mapping of percolation to the SIR model is straight forward: edges are occupied with probability $T_p$, equal to the time-integrated probability $T$ that an infection occurs on the edges. Here $T=1-e^{-\beta t}$ is the probability that a neighbor of an infected node is infected before it recovers, where $\beta$ is the disease-causing infection probability and $t$ is the time the infected node remains infective. If using the discrete time rather than continuous, which is common in computer simulation, then $T=1-(1-\beta)^{t}$, where $t$ is measured in time steps~\cite{newman2002}. The giant component appearing in the percolation process corresponds to the potential epidemic outbreak of disease with a non-negligible fraction of the network size.

To map the coevolving dynamics on multiplex network into percolation process, we need the following assumption that the information spreading is much faster that the epidemic spreading~\cite{newman2011}. This is reasonable for the Internet time, as information is easily transmitted worldwide through online social media, telephone, mass media, et al. Like the COVID-19 epidemics, the whole world gets to know its information soon after it outbreaks in Wuhan, China.
In addition, the vaccination in layer B can be regarded as a type of "disease" because each node in layer B can be in either infected states or vaccinated states~\cite{wang2014}. The disease spreading and vaccination are then viewed as two competing "diseases" on layer B. In the limit of large network size N, when two competing diseases spreads, it can be considered as if they were spreading non-concurrently, one after the other~\cite{newman2011}. So we can treat our coevolving dynamics as a fast dynamics of information spreading spreads first and a slow dynamics of disease spreading spreads subsequently.
\subsection{Quenching the fast dynamics}
First we consider the fast dynamics, i.e. the information spreading in layer A. This is a simple SIR process. Let $H^A_{i\to j}$ be the probability that node $i$ is not connected to the giant component via node $j$. Then $H^A_{i\to j}$ can be obtained by the self-consistency equations~\cite{brian2014}
\begin{equation}
H^A_{i\to j}=1-T^A+T^A\prod_{k\in \partial j\setminus i}H^A_{j\to k},
\end{equation}
where $T^A$ is the edge occupation probability in percolation, and $\partial j\setminus i$ is the neighbors of node $j$ except $i$. This equation represents that either the edge connecting $i$ and $j$ is not occupied, or although it is occupied, $j$ is not connected to the giant component through any of its neighbors other than $i$.
The probability that $i$ in the giant component is
\begin{equation}
P^A_i=1-\prod_{j\in \partial i} H^A_{i\to j},
\end{equation}
where $\partial i$ is the neighbor set of node $i$.
Mapping to SIR dynamics, $T^A=1-\mathrm{e}^{-\beta_A}$ when $t=1$. $H^A_{i\to j}$ is the probability that node $j$ by following the link from $i$ does not trigger out outbreaks with the transmissibility $T^A$. $P^A_i$ is the probability that a node $i$ in layer A triggers the epidemic outbreak in terms of $H^A_{i\to j}$.
\subsection{Fast dynamics as an outer field to slow dynamics}
Now consider the slow dynamics in layer $B$. Let $H^B_{i\to j}$ be the probability that node $i$ is not connected to the giant component via node $j$. This can happen because of (a) the node $j$ is vaccinated with probability $\lambda_{AB}P^A_i$; (b) the node $j$ is not vaccinated, but the edge $i\to j$ is not occupied with probability $1-\beta_B$; and (c) the node $j$ is not vaccinated and the edge $i\to j$ is occupied, but is not connected to the giant component via any of the neighbors. Conclude the above scenarios we have
\begin{equation}
\begin{split}
H^B_{i\to j}=&\lambda_{AB}P_{i}^A+\left(1-\lambda_{AB} P^A_i\right)\left(1-T^B\right)\\
&+T^B\left(1-\lambda_{AB}P^A_i\right)\prod_{k\in \partial j\setminus i}H^B_{j\to k},
\end{split}
\end{equation}
where $T^B=\beta_B$ because we simulate the SIR dynamics in discrete time and the infected nodes recover after one time step, corresponding to $t=1$ in the above mentioned definition. The probability that $i$ in the giant component is
\begin{equation}
P^B_i=1-\prod_{j\in \partial i} H^B_{i\to j}.
\end{equation}
This $P^B_i$ is the probability that a node $i$ in layer B triggers an epidemic outbreak in terms of $H^B_{i\to j}$.

Then according to ref.~\cite{min2018}, the size of epidemic when it originates from a seed $i$ is
\begin{equation}
S^B_i=\frac{1}{N}(1+\sum_{j=1,j\neq i}^NP^B_j).
\end{equation}
As the seed $i$ must be included in the epidemic outbreak, it corresponds to $1$ in the summation.
After obtaining the probability and the outbreak size, the average prevalence when the epidemic is initiated by a seed $i$ is defined as
\begin{equation}
\rho^P_i=P^B_i*S^B_i.
\end{equation}
Thus we can take $\rho^P$ as the indicator of node influence in layer B in the asymmetrically interacting processes on multiplex network. The calculation of influence is described in algorithm~\ref{alg_rank}.
\begin{algorithm}[h]
\caption{Influence Calculation in Multiplex Network}
\label{alg_rank}
\begin{algorithmic}[1]
\Require
the network of layer A $G_A$ and layer B $G_B$;
the information transmission rate $\lambda_A$, the disease transmission rate $\lambda_B$ and the vaccination rate $\lambda_{AB}$
\Ensure
the set of influence of each node i as the initial seed $\rho_i^P$
\State calculate $H^A_{i\rightarrow j}$ and $P_i^A$ from network A
\State calculate $H^B_{i\rightarrow j}$ and $P_i^B$ from network B based on the obtained value of $P_i^A$
\State calculate $S_i^B$ from network B based on the obtained value of $P_i^B$
\State calculate $\rho_i^P$ based on the obtained value of $P_i^B$ and $S_i^B$
\end{algorithmic}
\end{algorithm}
\begin{figure*}[htbp]
\begin{center}
\includegraphics[width=15 cm]{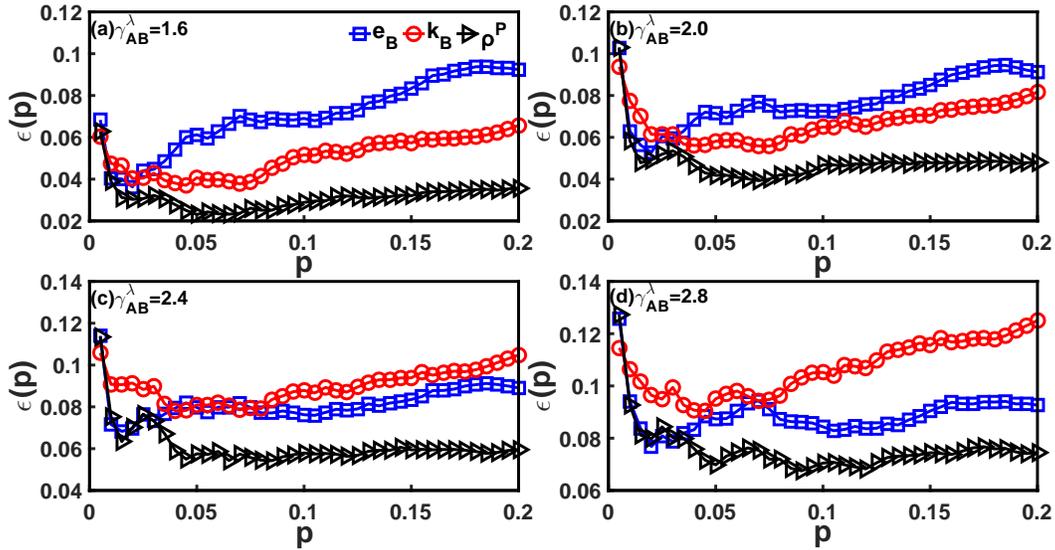}
\caption{The imprecision of $e_B$, $k_B$ and $\rho^P$ as a function of $p$ for different $\gamma^{\lambda}_{AB}$. The other parameters are set as: $\lambda_{AB}=0.7$, $\lambda_{BA}=0.1$, $m_s=0.1$.}
\label{figure5}
\end{center}
\end{figure*}
\section{Evaluation of the proposed method in identifying influential spreaders on multiplex network}
\begin{figure*}[htbp]
\centering
\includegraphics[width=15 cm]{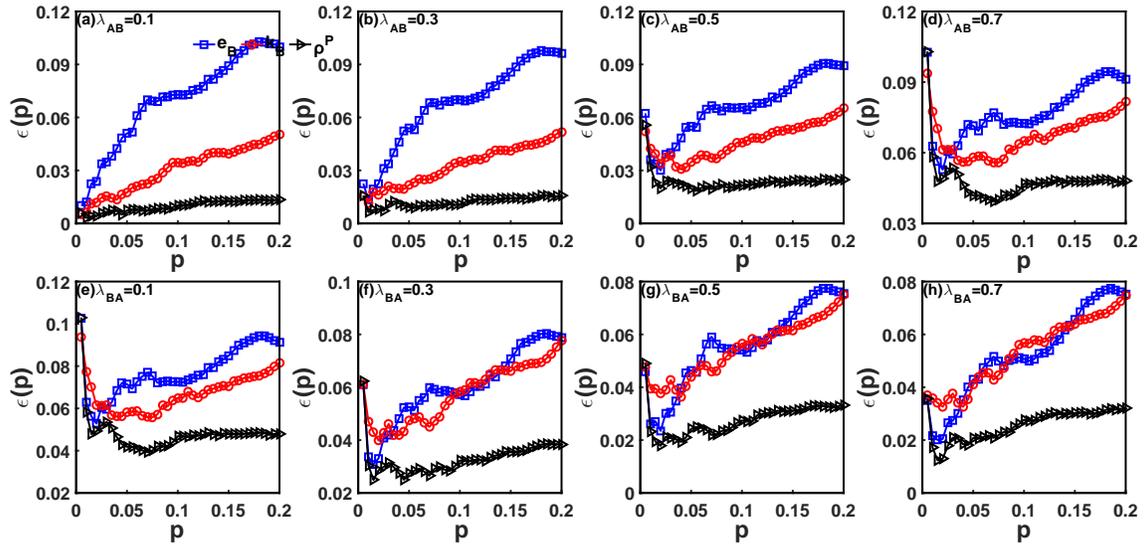}
\caption{The imprecision of $e_B$, $k_B$ and $\rho^P$ as a function of $p$ for different $\lambda_{AB}$ and $\lambda_{BA}$ respectively. The other parameters are set as: $\gamma^{\lambda}_{AB}=2.0$, $m_s=0.1$, $\lambda_{BA}=0.1$ in (a)-(d), $\lambda_{AB}=0.7$ in (e)-(h).}
\label{figure6}
\end{figure*}
Now we evaluate the proposed method in identifying influential spreaders on multiplex network. As we have revealed that the dynamical interplay of the two layers impacts the spreading influence of nodes, we compare the accuracy of $\rho^P$ with degree and eigenvector centrality under different values of the parameters. 

In Fig.~\ref{figure5}, we vary the two-layer relative spreading speed $\gamma^{\lambda}_{AB}$ and set all other parameters as fixed. It can be seen that the imprecision defined in 4.1 of $\rho^P$ is much lower than that of the degree and eigenvector centrality. Although at $p=0.005$ the imprecision of $\rho^P$ is a little bit higher than that of $k_B$, in most cases, $\rho^P$ is the best.

In Fig.~\ref{figure6} imprecisions under different $\lambda_{AB}$ and $\lambda_{BA}$ are demonstrated. It can be seen that, under all values of $\lambda_{AB}$ and $\lambda_{BA}$ respectively, the imprecision of $\rho^P$ is significantly lower than that of $k_B$ and $e_B$ in almost all the cases.
\begin{figure*}[htbp]
\begin{center}
\includegraphics[width=15 cm]{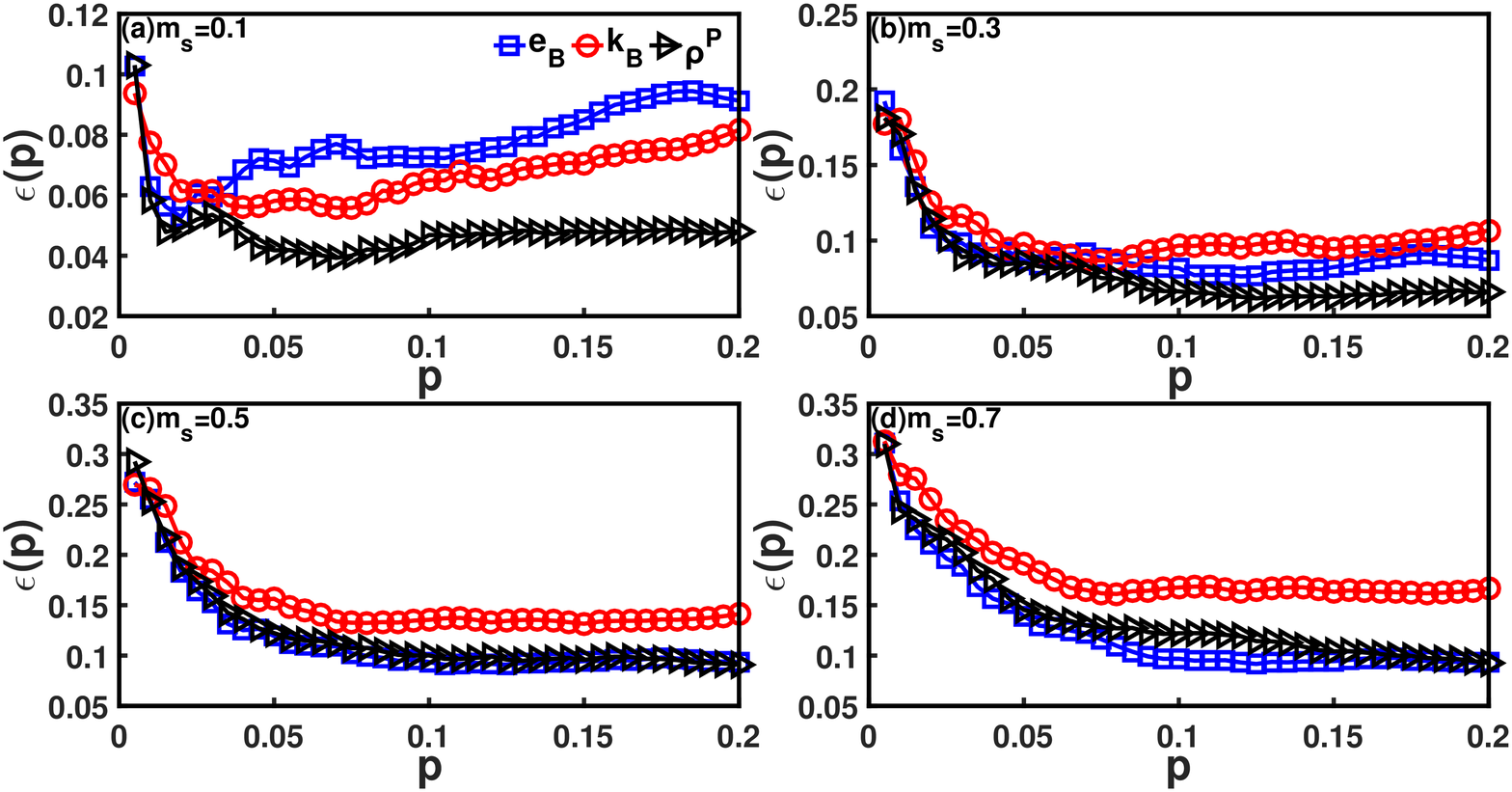}
\caption{The imprecision of $e_B$, $k_B$ and $\rho^P$ as a function of $p$ for different $m_s$. The other parameters are set as: $\gamma^{\lambda}_{AB}=2.0$, $\lambda_{AB}=0.7$, $\lambda_{BA}=0.1$.}
\label{figure7}
\end{center}
\end{figure*}

\begin{figure}[htbp]
\begin{center}
\includegraphics[width=9.1 cm]{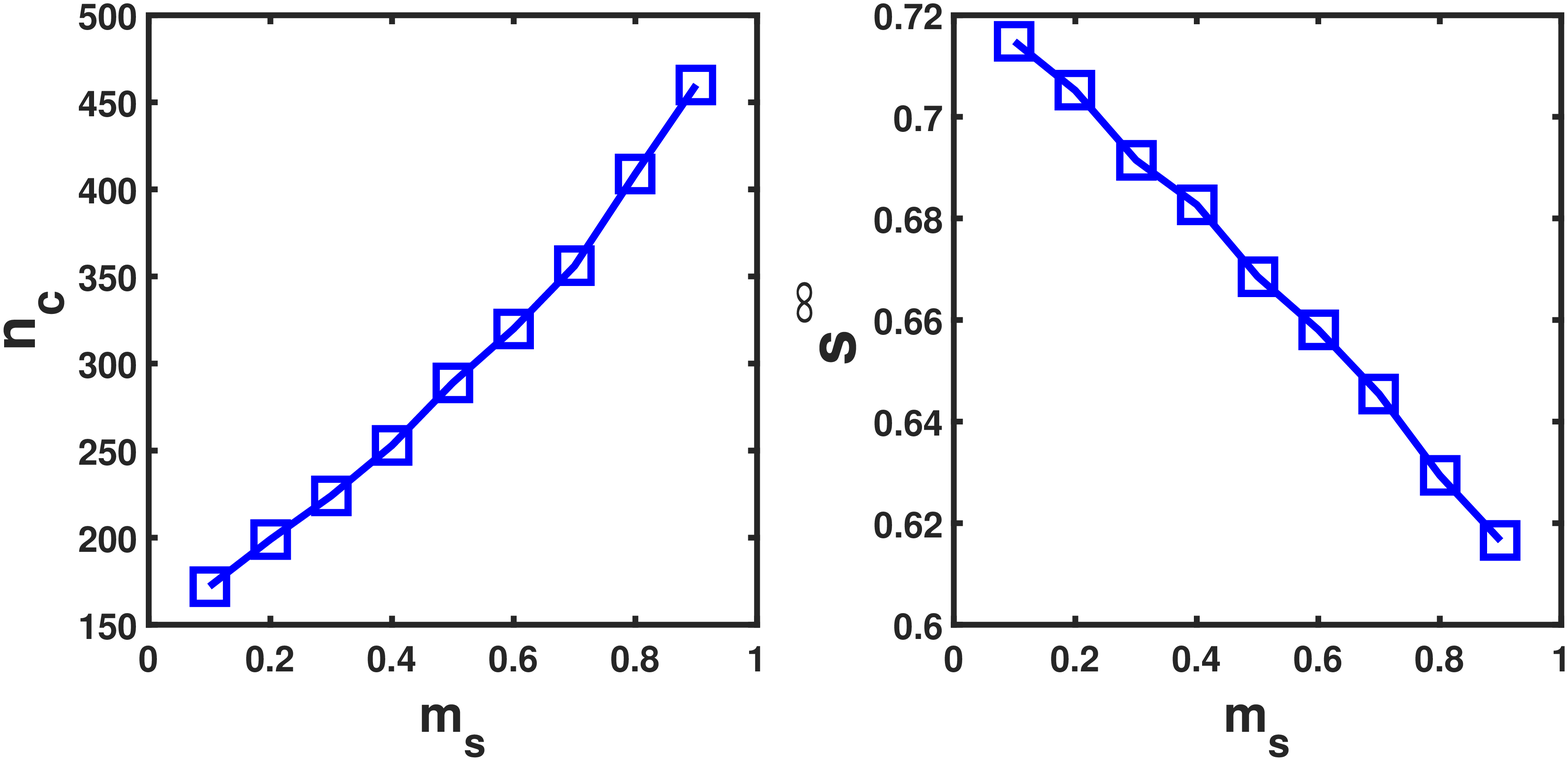}
\caption{The number of components $n_c$ and the size of giant component as a function of degree correlation $m_s$. The other parameters are set as: $\gamma^{\lambda}_{AB}=2.0$, $\lambda_{AB}=0.7$, $\lambda_{BA}=0.1$.}
\label{figure8}
\end{center}
\end{figure}

Finally, we demonstrate the imprecisions of the three methods under different degree correlations $m_s$. From Fig.~\ref{figure7} (a) and (b), we can see that the $\rho^P$ is the best indicator for spreading influence. When the degree correlation increases to 0.5 or 0.7, the imprecision of $\rho$ is equal to or slightly higher than that of $e_B$.  We think this is because when we calculate the $\rho^P$, it is based on the assumption that the layer A spreads information first. When the degree correlation is large, the hub nodes in layer B is probably to be vaccinated. Then layer B is separated by the vaccinated nodes into several small clusters and a giant component. The calculation of $\rho^P$ on layer B is less accurate because the network has been dismantled. To evaluate this, we calculate the number of components and the size of giant component when $m_s$ varies. As shown in Fig.~\ref{figure8}, with the increase of degree correlation $m_s$ the number of components in layer B increases and the size of the giant component decreases.
In general, $\rho^P$ is the best way to predict the spreading influence of nodes under asymmetrically interacting processes. These results imply that in the modern world, where the spreading of information is very easy and fast to world-wide, the control strategy based on the contact layer alone will loss its effectiveness. To accurately identify the most influential spreaders and control the spread of epidemic disease, we need not only the physical contact network data, but also the information transmission network data and the coupling parameters.
\section{Conclusion}
In this paper, we study on how the two-layer coupling factors impact the centrality of nodes to predict their spreading influence and propose a method to identify the most influential spreaders in multiplex networks. The results show that the benchmark centralities like degree and eigenvector centrality in one layer alone can not predict the influential spreaders accurately due to the interplay between the spreading of information and disease on multiplex networks. The relative spreading speed of two layers, the interlayer dynamical coupling strength and the two-layer degree correlation play an important role in affecting the spreading influence of nodes. By mapping the coevolving spreading dynamics into bond percolation, we use the message-passing approach to calculate the epidemic outbreak size when spreading is initiated by a single seed. The obtained measure takes both the intralayer and interlayer structural and dynamical information into account and is thus very accurate in identifying the most influential disease spreaders. Our work provides new ideas for making effective epidemic control strategy and gives a feasible framework to study the identification of critical nodes on multilayer networks. Although we study on the identification of critical nodes in the asymmetrically interacting spreading dynamics, the interacting dynamics on the multilayer network can be interdependent or competitive. How these interacting dynamics influences the functional importance of nodes is the question in future study.

\ifCLASSOPTIONcompsoc
  \section*{Acknowledgments}
\else
  \section*{Acknowledgment}
\fi

This work is supported by the National Natural Science Foundation of China (No. 61802321, 11975099), the Sichuan Science and Technology Program (No. 2020YJ0125), and the Natural Science Foundation of Shanghai (No. 18ZR1412200)

\ifCLASSOPTIONcaptionsoff
  \newpage
\fi

\end{document}